\documentclass[5p,number]{elsarticle}
\usepackage{color}
\usepackage{graphicx}
\usepackage{psfrag}
\usepackage{hyperref}
\usepackage{amsmath,amssymb}

\newcommand{\cp}[1]{\mathbb{C}\mathrm{P}^{#1}}
\newcommand{\ii}{\mathrm{i}}
\newcommand{\abs}[1]{\left\vert #1 \right\vert}
\newcommand{\abstx}[1]{\vert #1 \vert}
\newcommand{\cpn}{$\mathbb C\mathrm P^n$}
\newcommand{\C}{\mathbb C}
\newcommand{\iu}{\mathrm i}
\DeclareMathOperator{\tr}{tr}
\DeclareMathOperator{\ee}{e}
\DeclareMathOperator{\diag}{diag}

\journal{arXiv}
\submitdate{February 13, 2009}
\begin{document}
\title{Instanton constituents and fermionic zero modes in twisted {\cpn} models}

\author[erl]{Wieland Brendel}
\ead{wieland@theorie3.physik.uni-erlangen.de}
\author[erl,reg]{Falk Bruckmann}
\ead{falk.bruckmann@physik.uni-regensburg.de}
\author[jen]{Lukas Janssen}
\ead{lukas.janssen@uni-jena.de}
\author[jen]{Andreas Wipf}
\ead{wipf@tpi.uni-jena.de}
\author[jen]{Christian Wozar}
\ead{christian.wozar@uni-jena.de}
\address[erl]{Institut f{\"u}r Theoretische Physik III,
 Universit{\"a}t Erlangen-N{\"u}rnberg, D-91058 Erlangen, Germany}
\address[reg]{Institut f{\"u}r Theoretische Physik, Universit{\"a}t Regensburg, D-93040 Regensburg, Germany}
\address[jen]{Theoretisch-Physikalisches Institut,
Universit{\"a}t Jena, D-07743
Jena, Germany}

\begin{abstract}
\noindent
We construct twisted instanton solutions of {\cpn} models. 
Generically a charge-$k$ instanton splits into $k(n+1)$ well-separated and 
almost static constituents carrying fractional topological charges and being 
ordered along the noncompact direction. The locations, sizes and charges of the 
constituents are related to the moduli parameters of the instantons. We sketch 
how solutions with fractional total charge can be obtained. We also calculate 
the fermionic zero modes with quasi-periodic boundary conditions in the 
background of twisted instantons for minimally and supersymmetrically coupled 
fermions. The zero modes are tracers for the constituents and show a 
characteristic hopping. The analytical findings are compared to results 
extracted from Monte-Carlo generated and cooled configurations of the 
corresponding lattice models. Analytical and numerical results are in full 
agreement and it is demonstrated that the fermionic zero modes are excellent 
filters for constituents hidden in fluctuating lattice configurations.
\end{abstract}

\maketitle

\graphicspath{{figures/}}

\section{Introduction}
\noindent
Nonlinear sigma models in two dimensions have long been used as testing ground 
for strongly coupled gauge theories \cite{Polyakov:1975rr}. They are scale 
invariant on the classical level and asymptotically free at the quantum level. 
The ubiquitous {\cpn} models possess regular instanton solutions, the 
topological charges of which yield lower BPS-bounds on the action, they have a 
chiral anomaly when coupled to fermions, generate a dynamical mass by 
non-perturbative effects at zero temperature and a thermal mass $\propto g^2 T$ 
at finite temperature. They have numerous interesting applications to condensed 
matter physics (for a review see \cite{Tsvelik:1995}) and also have been used 
to study the sphaleron induced fermion-number violation at high temperature 
\cite{Mottola:1988ff}.

In the present work we consider {\cpn} models at finite temperature, i.e., 
\emph{Euclidean models} with imaginary time having period 
$\beta=1/k_\text{B}T$. These models possess instanton solutions with finite 
action and the dimension of the moduli space in a given instanton sector 
depends on the topology of the Euclidean space-time. For example, on the 
two-torus the charge-$k$ instantons of {\cpn} depend on as many collective 
parameters as the instantons of $\cp{n+1}$ with one charge less  
\cite{Aguado:2001xg}. In the present work we do not compactify space such that 
space-time is a cylinder.

For suitable field variables the selfduality equation for {\cpn} instantons 
reduces to Cauchy-Riemann conditions such that all instantons are known 
explicitly for the plane, cylinder and to\-rus. On the plane they are given by 
rational functions of the complex coordinate $z$ and on the cylinder by 
suitable periodic generalizations thereof, see below.

In a previous work \cite{Bruckmann:2007zh} one of us introduced the twisted 
$O(3)$ model (which is equivalent to the $\cp{1}$ model) and showed that 
generically the unit charged instantons in this model dissociate into 
two fractional charged constituents, sometimes called `instanton quarks'. Again there is 
a close analogy to the corresponding situation in Yang-Mills theories, where 
instantons with nontrivial holonomy along the compact direction of a 
four-dimensional cylinder possess magnetic mo\-no\-po\-les as cons\-ti\-tuents 
\cite{Kraan:1998pm,Lee:1998bb,Bruckmann:2003yq,Bruckmann:2007ru}.

We extend the work in \cite{Bruckmann:2007zh} in several
directions. First we construct the $k(n+1)$ constituents of 
{\cpn} instantons with charge $k$ and twisted boundary conditions
and relate their positions, sizes and fractional charges to the
collective parameters of the instantons. Then we calculate and analyze the zero modes 
of the Dirac operator  for minimally coupled fermions with quasi-periodic 
boundary conditions in the background of the twisted instantons. 
We show that the zero modes can be used as tracers for the instanton 
constituents: they are localized to the latter, to which constituent depends on the boundary condition.
Again, this has close analogies in four-dimensional Yang-Mills theories with $1$ (or $2$) compact dimensions 
\cite{GarciaPerez:1999ux,Bruckmann:2003ag,Ford:2005sq}.

{\cpn} spaces admit a K\"ahler structure such that the two-dimensional {\cpn} 
models admit a supersymmetric extension with two supersymmetries. These models 
contain $4$-fermi interactions and the Dirac operator is given by the linearized 
field equation for the fermions. We calculate the zero modes of this operator. 
There exists always one zero mode with squared amplitude being proportional to 
the action density of the twisted instanton.

We supplement our analytic studies by numerical simulations. With known 
algorithms we produce typical field configurations for various {\cpn} models 
with twisted boundary conditions. Then we apply standard lattice cooling 
techniques to extract the instantons and their constituents from a given 
(thermalized) configuration. Again we find that an instanton of charge $k$ 
consists of $k(n+1)$ constituents. How the corresponding instanton constituents 
in Yang-Mills theories emerge in the process of cooling/smearing has been 
studied in 
\cite{GarciaPerez:1999hs,ilgenfritz:02a,bruckmann:04b,ilgenfritz:04a,ilgenfritz:05,Bornyakov:2008im}. 
Next we compute and analyze the zero modes of the overlap Dirac operator. We 
find good agreement between our numerical and analytical results.

Our results demonstrate that {\cpn} models and Yang-Mills theories share one 
more common feature: in both models the twisted instantons generically split 
into well-separated and almost static constituents and in both models the 
fermionic zero modes trace these constituents. In $SU(N)$ Yang-Mills theories the 
number of constituents is given by $k$ and the rank  
whereas for {\cpn} models it is given by $k$ and by $n$. On the lattice the 
fermionic zero modes are excellent filters for the constituents. Even without 
much cooling the zero modes detect the constituents of the fully cooled 
configurations.


\section[The CP(n) model in the continuum and on the lattice]
{The {\cpn} model in the continuum and on the lattice}
\noindent
The two-dimensional {\cpn} model \cite{Eichenherr:1978qa, DAdda:1978uc} can be 
formulated in terms of a complex $(n+1)$-vector $u=(u_0,\dots,u_n)^T$ subject 
to the constraint $u^\dagger u =1$. The Euclidean action is given by
\begin{equation}
S = \frac{2}{g^2} \int \mathrm d^2x\,\left(D_\mu u\right)^\dagger D_\mu u,\quad 
D_\mu = \partial_\mu - \iu A_\mu.
\end{equation}
It is invariant under local $U(1)$ gauge transformations
\begin{equation}
u_j(x) \mapsto \ee ^ {\ii \lambda(x)} u_j(x),\quad A_\mu(x)\mapsto 
A_\mu(x)+\partial_\mu\lambda(x),
\end{equation} 
as well as global transformations
\begin{equation} 
u_{j}(x) \mapsto \mathcal U_{jl} u_{l}(x)
 \end{equation}
with a constant matrix $\mathcal U\in U(n+1)$. The gauge field $A_\mu$ can be 
eliminated from the action by using its algebraic equation of motion,
\begin{equation}
A_\mu = - \iu u^\dagger \partial_\mu u.
\end{equation}
The integer-valued instanton number,
\begin{equation}
Q = \int \mathrm d^2x\, q(x) \quad \text{with} \quad q(x) = 
\frac{1}{2\pi}\epsilon_{\mu\nu}\partial_\mu A_\nu(x),
\end{equation}
can be interpreted as the quantized magnetic flux in a fictitious third 
dimension. At infinity $u$ must approach a pure gauge, $u(x)\rightarrow 
\ee^{\iu\lambda(x)}c$ and $Q$ is just the winding number of the map 
$x\rightarrow\ee^{\iu\lambda(x)}$ at infinity, an element of the first 
homotopy group of $U(1)$.

Configurations minimizing the action in $S\geq 4\pi \abstx{Q}/g^2$ are called 
\emph{instantons}. They fulfill first order \emph{self-duality equations}. The 
most general instanton solution with instanton number $Q=k\in\mathbb N$ can be 
written in homogeneous coordinates $v_j$ as
\begin{equation}
\label{eqn:hom_coord}
u_j(x) = \frac{v_j(z)}{\abs{v(z)}}, \quad j=0,\dots,n,
\end{equation}
with $\{v_j\}$ a set of polynomials of the complex coordinate $z = x_1+\iu x_2$ 
with no common root and maximum degree $k$. The topological charge density of 
an instanton configuration then reads
\begin{equation}
\label{eq:qx}
q(x) = \frac{1}{4\pi}\Delta \ln\abs{v(z)}^2.
\end{equation}


\subsection*{Lattice formulation}
\noindent
For the bosonic model the lattice regularization can be obtained as described 
in \cite{DiVecchia:1981eh,Wolff:1992ze}. After introducing the matrix-valued 
gauge invariant field
\begin{equation} 
P(x) = u(x)u^\dagger(x),
\end{equation}
which projects onto the one-dimensional subspace spanned by $u$, one finds
\begin{equation}
\tr \left[\partial_\mu P ~ \partial_\mu P\right] = 2\partial_\mu 
u^\dagger\partial_\mu u+ 2 \left(u^\dagger\partial_\mu u\right)^2 =g^2\mathcal 
L.
\end{equation}
This equation, valid for the model defined on a continuous space-time, is 
discretized naively with the forward derivative, $\partial_\mu P\mapsto 
P_{x+\hat{\mu}}-P_x$, such that
\begin{equation}
\tr \left[\partial_\mu P ~ \partial_\mu P\right] \mapsto 2d - 2\sum_\mu \tr 
\left[P_x P_{x+\hat{\mu}}\right].
\end{equation}
Therefore, the action, up to an irrelevant additive constant, takes the form
\begin{equation}
S = -\frac{2}{g^2}\sum_{x,\mu} \tr\left[P_xP_{x+\hat{\mu}}\right] = 
-\frac{2}{g^2}\sum_{x,\mu} \abs{u_x^\dagger u_{x+\hat\mu}}^2.
\end{equation}
The simulations of the lattice models have been performed with the help of an 
overrelaxation algorithm \cite{Wolff:1992ze}. In addition, to investigate the 
topological properties, we cooled the lattice configurations 
\cite{Berg:1981nw}. For a given configuration one \emph{cooling step} consists 
of minimizing the action locally on a randomly chosen site $x$. This is 
achieved by constructing $Q_x=\sum_\mu (P_{x+\hat\mu}+P_{x-\hat\mu})$ and 
replacing $u_x$ by the eigenvector corresponding to the largest eigenvalue of 
$Q_x$. A \emph{cooling sweep} corresponds to one cooling step per lattice site 
on average. Using this procedure the instanton constituents naturally emerge 
from the locally fluctuating fields.

For the topological charge on the lattice we used the geometric definition in 
\cite{Berg:1981er} leading to an integer-valued instanton number. This 
definition and the chosen lattice action are sufficient for the analysis of 
global topological properties in the vicinity of classical configurations. 
Thus, we are not affected by the improper scaling behavior of the dynamical 
$\cp{n}$ models with $n\le 2$ \cite{Luscher:1981tq}.


\section{Instantons at finite temperature}\label{sec:instantons}

\begin{figure*}[tb]
\centering
\includegraphics{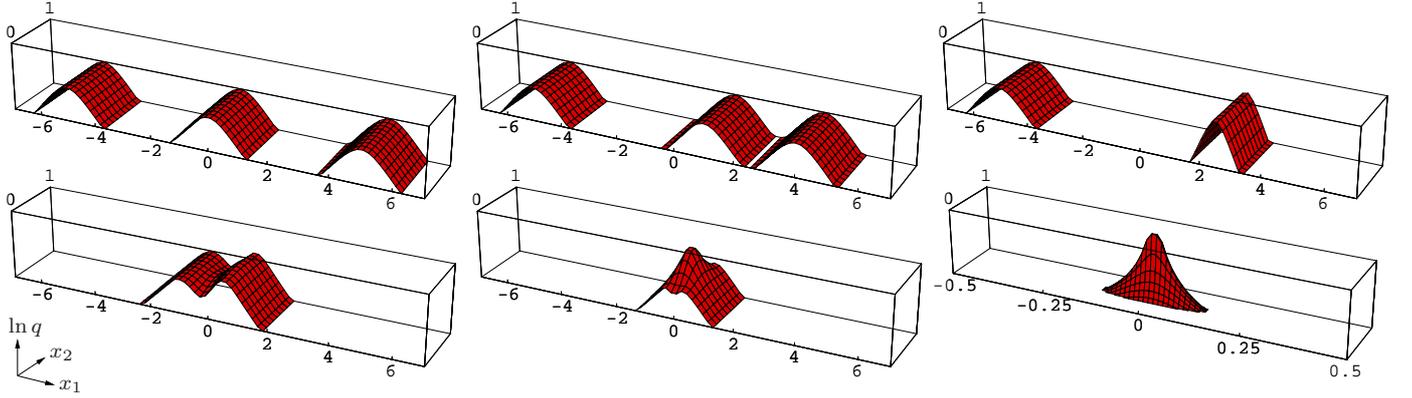}
\caption{(Color online.) Logarithm of the topological density for the 
$1$-instanton solution of  the $\cp{2}$ model (see \eqref{eq:qx} 
and \eqref{eq:abs_w2__1-instanton}) with symmetric constituents, 
$\mu_1=\mu_2-\mu_1=1-\mu_2=1/3$  (cut off below $\ee^{-5}$). The 
parameters $\lambda_i$ are chosen such that the constituents are localized 
according to \eqref{eqn:const_positions} from left to right at $(a_1, a_2, a_3) 
= (-5,0,5)$, $(-5,1,4)$, $(-5,7,-2)$ (first line) and $(-1,0.5,0.5)$, 
$(0,0,0)$, $(3,-1,-2)$ (second line). Note that the $x_1$-range has been 
changed in the lower right panel.}\label{fig:density_1-instanton_cp2}
\end{figure*}

\noindent
For a quantum system at inverse temperature $\beta$ we identify $z \sim z + 
\iu\beta$. Since $\beta$ is the only length scale in the problem we measure all 
lengths in units of $\beta$. In particular the coordinates become dimensionless 
and we identify $z\sim z+\iu$. Periodic $k$-instanton solutions (`calorons') 
are given by
\cite{Affleck:1980ij, Actor:1985yh}
\begin{equation}
v_\text{per}(z) = b^{(0)} + b^{(1)} \ee^{2\pi z} + \cdots + b^{(k)} \ee ^{2\pi 
k z}.\label{pinst}
\end{equation}
By a global $U(n+1)$ symmetry transformation one can rotate $v_\text{per}$ such 
that the constant (and per assumption non-vanishing) vector 
$b^{(k)}\in\C^{n+1}$ points in the $0$-direction, $b^{(k)}_j = 
b^{(k)}_0\delta_{j0}$.

The \emph{twisted model} is only quasi-periodic in the imaginary time 
direction. This means that the components $v_j$ of $v$ are periodic up to 
phases  $\ee^{2\pi\iu\mu_j}$ with $\mu_j\in[0,1)$, i.e., the 
vectors $v$ and $u$ are periodic up to a diagonal element of the global 
symmetry $U(n+1)$. The $U(n+1)$-invariants like $\abstx{v}$ and $A_\mu$ and 
hence also $q$ stay periodic. Without loss of generality we assume the phases 
to be ordered according 
to~$\mu_0\!\leq\!\mu_1\!\leq\!\dots\!\leq\!\mu_n$.

For the general solutions of the twisted model we consider the  Fourier ansatz
\begin{equation}
v_j(z)=\ee^{2\pi \mu_j z}\sum_{s=-\infty}^{\infty} b^{(s)}_j \ee^{2\pi s z}
\end{equation}
and demand the coefficients $b^{(s)}_j $ to be non-vanishing only for a finite 
range of $s$ (for each component $j$). This is because the corresponding 
maximum and minimum of the powers
\begin{equation}
\kappa_\text{max}=\max_{j,s\colon b^{(s)}_j\neq 0} (s+\mu_j), \quad 
\kappa_\text{min}=\min_{j,s\colon b^{(s)}_j\neq 0} (s+\mu_j)
\end{equation}
then yield a finite topological charge $Q$. According to \eqref{eq:qx} one has 
to compute the following surface integrals
\begin{equation}\begin{split}
Q & =\frac{1}{4\pi}\int_0^\beta \mathrm dx_2 \,\partial_1 \ln 
\abs{v}^2\Big|^{x_1\to\infty}_{x_1\to - \infty}\\
& = \kappa_\text{max}-\kappa_\text{min}\\
&\in \mathbb N_0+\left\{\mu_j-\mu_l\, \vert \, j,l=0,\dots,n\right\}.
\end{split}\end{equation}
Hence the total topological charge in the twisted model can have a fractional 
part, whose values are restricted by the boundary conditions. By a global 
transformation we enforce $\kappa_\text{min}$ to be taken on in the $0$th 
component and by a (non-periodic) local transformation we further set 
$\mu_0=0$ and $\kappa_\text{min}=0$, such that $Q=\kappa_\text{max}$. 
According to Eq.~\eqref{eqn:hom_coord}, these powers also govern the asymptotic 
values of the fundamental fields $u_j$.

In the following we will mainly analyze \emph{twisted instantons} with 
integer-valued instanton number $Q=k\in \mathbb N$. They are obtained by 
$\kappa_\text{max}=k$ taken on in the $0$th component, i.e., the highest 
coefficient $b^{(k)}$ points in the $0$-direction, $b^{(k)}_j = 
b^{(k)}_0\delta_{j0}$. Thus one can obtain the components $v_j$ by multiplying 
each component $v_{\text{per},j}$ from \eqref{pinst} with $\exp(2\pi \mu_j 
z)$, which yields
\begin{equation}
\label{eq:general_twisted_instanton_solution}
v(z)=\Omega \, v_\text{per}(z),\quad \Omega=\diag\left(\ee^{2\pi 
\mu_0z},\ldots,\ee^{2\pi\mu_n z}\right).
\end{equation}

For $n=1$ the known twisted unit charged instanton solution 
\cite{Bruckmann:2007zh} can be recovered in terms of the gauge invariant field
\begin{equation}
\frac{v_1(z)}{v_0(z)} = \frac{b^{(0)}_1 \ee^{2\pi \omega z}}{b_0^{(0)} + 
b_0^{(1)} \ee ^{2\pi z}}.
\end{equation}
We made use of $\mu_0=0$ and $b^{(1)}_1=0$ and denoted $\mu_1$ by 
$\omega$.


\subsection{One-instanton sector}
\noindent
In order to explore the topological density of the instantons we first consider 
solutions with unit charge $Q=k=1$. We multiply $v$ by a constant such that 
$b^{(0)}_0=1$ and afterwards shift the Euclidean time $x_2$ such that  
$b^{(1)}_0$ becomes real and non-negative. For this choices the density 
$\abstx{v}^2$ only depends on the absolute values $\lambda_j=\abstx{b^{(0)}_j}$ 
with $j=0,1,\dots,n$. If, in addition, we define $\lambda_{n+1}= \abstx{b^{(1)}_0}$ 
and $\mu_{n+1}=1$, then it can be written in the condensed form
\begin{equation}\label{eq:symmversion}
\abs{v(z)}^2 = \sum_{i=0}^{n+1} \lambda_i^2 \ee^{4\pi\mu_i x_1}+ 
2\lambda_{n+1} \ee^{2\pi x_1}\cos(2\pi x_2).
\end{equation}
The corresponding topological charge density splits into $n+1$ constituents at 
most. For $\cp{2}$ this is illustrated in 
Fig.~\ref{fig:density_1-instanton_cp2} which shows $\ln q(x)$ for various 
choices of the parameters $\lambda_i$.

\begin{figure}[p]
\centering\includegraphics{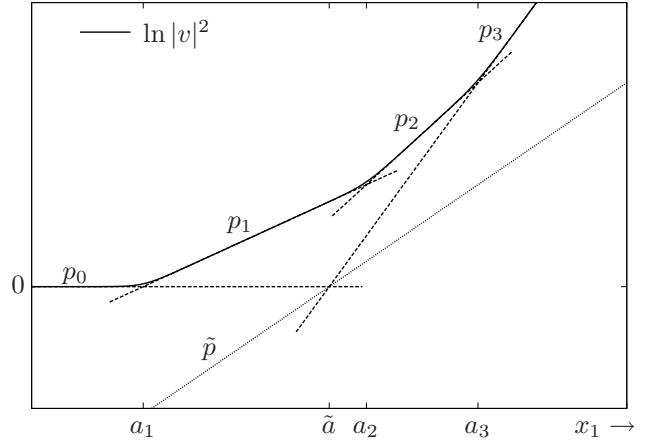}
\caption{$\ln 
\abstx{v}^2$ and exponents $p_i$ and $\tilde p$ as a function of $x_1$, see 
Eqs. \eqref{eq:abs_w2__1-instanton}--\eqref{eq:exponents_abs_w2__1-instanton}, 
in the $\cp{2}$ model for the case of $(a_1,a_2,a_3)=(-5,1,4)$, which leads to 
three well-separated constituents (equivalent to 2nd example in 
Fig.~\ref{fig:density_1-instanton_cp2}).}\label{fig:abs_v2_-5_1_4}
\end{figure}

\begin{figure}[p]
\centering\includegraphics{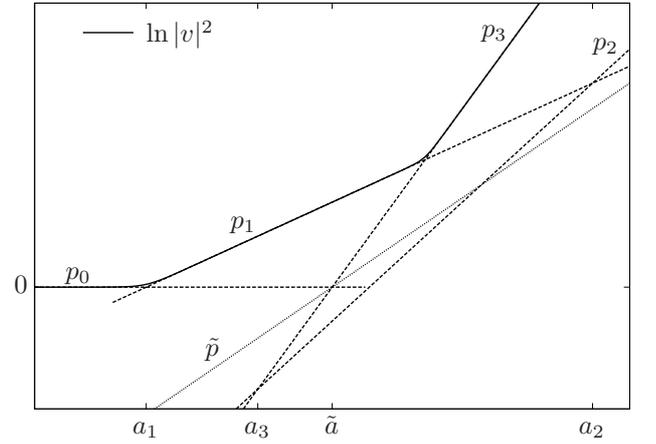}
\caption{$\ln \abstx{v}^2$ and exponents $p_i$ and $\tilde p$ as a function of 
$x_1$, for the case of  $(a_1,a_2,a_3)=(-5,7,-2)$, where the second and third 
constituent merged (equivalent to 3rd example in 
Fig.~\ref{fig:density_1-instanton_cp2}).}\label{fig:abs_v2_-5_7_-2}
\end{figure}

\begin{figure}[p]
\centering\includegraphics{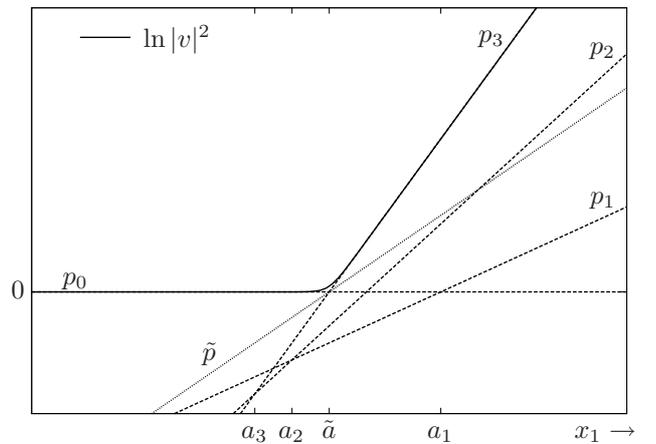}
\caption{$\ln \abstx{v}^2$ and exponents $p_i$ and $\tilde p$ as a function of
$x_1$, for the case of  $(a_1,a_2,a_3)=(3,-1,-2)$, where the time-dependent
$\tilde p$-term becomes relevant (equivalent to 6th example in
Fig.~\ref{fig:density_1-instanton_cp2}).}\label{fig:abs_v2_3_-1_-2}
\end{figure}

For the general {\cpn} models the occurrence of the constituents can be 
understood geometrically. To see this more clearly we write
\begin{equation}\label{eq:abs_w2__1-instanton}
\abs{v(z)}^2 = \sum_{i=0}^{n+1} \ee^{\,p_i(x_1)} + 2 \ee^{\,\tilde p(x_1)} 
\cos(2\pi x_2),
\end{equation}
with
\begin{equation}\label{eq:exponents_abs_w2__1-instanton}\begin{split}
p_i(x_1) \!& =4\pi\mu_i x_1 + 2\ln \lambda_i,  \\
\tilde p(x_1)  & = 2\pi x_1 + \ln \lambda_{n+1}.
\end{split}\end{equation}
In particular
\begin{equation}\begin{split}
p_0(x_1)\!&=0, \\
p_{n+1}(x_1)\!&=4\pi x_1+2\ln\lambda_{n+1}=2\tilde p(x_1).
\end{split}\end{equation}
We compare the graphs of these $n+3$ linear functions, see 
Figs.~\ref{fig:abs_v2_-5_1_4}--\ref{fig:abs_v2_3_-1_-2} for three examples in 
the $\cp{2}$ model amounting to five exponential terms.

The dominant contribution to $\abstx{v}^2$ in (\ref{eq:abs_w2__1-instanton}) at 
a fixed point $x_1$ comes from the exponential term whose graph is above the 
lines defined by the other exponential terms. Hence $\ln \abstx{v}^2$ is 
piecewise linear in the direction $x_1$ up to exponentially small corrections 
that are maximal in transition regions, where the highest lying graphs 
intersect.


Note that for a strictly linear $\ln \abstx{v}^2$ the topological density $q 
\propto \Delta \ln \abstx{v}^2$ would vanish exactly, whereas at cusps 
generated by intersections of linear parts the topological density would be a 
Dirac delta distribution (in $3$+$1$ dimensional Yang-Mills theory a similar 
singular localization can be obtained in the far-field limit 
\cite{Bruckmann:2002vy,Bruckmann:2003ag}). As this is a good approximation to 
the actual $\ln \abstx{v}^2$, we conclude that the topological density of the 
twisted instantons \emph{splits into constituents} localized at the 
intersection points of the lines. Because of the ordering of the $\mu$'s, 
the slopes of the linear functions $p_i$ are ordered as well. Note that for 
$x_1<-1/(2\pi\mu_1) \ln \lambda_1$ the term $\exp(p_0)$ dominates such that 
$\ln\abstx{v}\approx 0$ on the left of all constituents. Correspondingly, 
$\ln\abstx{v}\approx 4\pi x_1$ on the right of all constituents.

We obtain the maximum number of constituents, if all consecutive graphs 
intersect separately and above the rest of the graphs, respectively. More 
precisely said, the twisted instanton of {\cpn} splits into $n+1$ constituents, 
if, and only if, $a_1\ll a_2\ll \dots\ll a_{n+1}$,\footnote{Thereby we do not 
want to take the condition $a_{i-1}\ll a_{i}$ too literally. It is sufficient, 
if $a_{i-1}$ is not close to $a_{i}$, whereas the required distance is 
determined by the slopes of $p_i-p_{i-1}$ and $p_{i-1}-p_{i-2}$.} whereas $a_i$ 
is the intersection point of the lines $p_{i-1}$  and $p_i$,
\begin{equation}
\label{eqn:const_positions}
a_i= - \frac{\ln \left(\lambda_i/\lambda_{i-1}\right)}{2\pi\left(\mu_i-\mu_{i-1}\right)}, \quad 
i=1,\dots,n+1,
\end{equation}
i.e., in particular the twist parameters $\mu_i$ are distinct, 
$0=\mu_0<\mu_1<\dots<\mu_n<\mu_{n+1}=1$. These positions $a_i$ are 
arbitrary provided the corresponding $\lambda$-parameters are chosen according 
to
\begin{equation}\label{eq:ln_lambda__a(j)}
\ln \lambda_i = - 2\pi\sum_{l=1}^i (\mu_l - \mu_{l-1}) a_l 
\quad\text{with}\quad \mu_0=0.
\end{equation}

Let us discuss the case of the well-separated constituents. In the neighborhood 
of the intersection point $a_i$ of the lines $p_{i-1}$ and $p_i$ we can 
approximate
\begin{equation}
\abs{v(z)}^2 \approx \lambda^2_{i-1} \ee^{4\pi\mu_{i-1} x_1} + \lambda_i^2 
\ee^{4\pi\mu_i x_1},
\end{equation}
for $x_1$ not too far from the constituent $i$,
\begin{equation}
\frac{1}{2}\left(a_{i-1}+a_i\right)\leq x_1\leq 
\frac{1}{2}\left(a_i+a_{i+1}\right), \quad i=1,\dots,n+1.
\end{equation}
For $i=1$ the lower bound for $x_1$ is $-\infty$  and for $i=n+1$ the upper 
bound is $+\infty$. The contribution to the topological density of the $i$th 
constituent is
\begin{equation}
q_{\text{const},i}(x)
\approx \frac{\pi 
\left(\mu_i-\mu_{i-1}\right)^2}{\cosh^2\left[2\pi(\mu_i-\mu_{i-1})\left(x_1 - 
a_i\right)\right]}.
\end{equation}
This shape is the same for all {\cpn} models, cf.\ \cite{Bruckmann:2007zh} for 
the  $\cp{1}$ case. The constituent decays exponentially with characteristic 
length $2\pi(\mu_i-\mu_{i-1})$ (measured in units of $\beta$) away from 
its position $a_i$. It has a fractional topological charge 
$Q_{\text{const},i}=\mu_i-\mu_{i-1}$ and these charges add up to $1$ as 
they should. In terms of the linear graphs the fractional topological charge is 
proportional to the difference of slopes of the lines that meet (which also 
would give the amplitude of the delta distribution mentioned above), and the 
total charge is the sum of all slope differences, which indeed bend the graph 
from $p_0$ with slope $0$ to  $p_{n+1}$ with slope $4\pi$ eventually.

Neighboring constituents can merge adding up the fractional topological 
charges. This can be understood as `pulling down' the line that connects the 
two constituents in the graph of $\ln\abstx{v}^2$, in other words by choosing 
the corresponding parameter $\lambda_i$ small (cf.\ 
Figs.~\ref{fig:abs_v2_-5_7_-2}--\ref{fig:abs_v2_3_-1_-2}).

Under which circumstances does the time-dependence of $\abstx{v}^2$ contained 
in the last term of Eq.~\ref{eq:abs_w2__1-instanton} (which is proportional to 
$\exp \tilde p(x_1)$) play a role? Since the three graphs of $p_0$, $p_{n+1}$ and 
$\tilde p$ intersect at the point $(\tilde a,0)$ with $2\pi\tilde a =-\ln \lambda_{n+1}$ 
we have
\begin{equation}
\tilde p(x_1)\leq \max\left\{p_0(x_1),p_{n+1}(x_1)\right\},
\end{equation}
such that the time-dependent $\tilde p$-term can contribute to the sum in 
\eqref{eq:abs_w2__1-instanton} only in the neighborhood of $\tilde a$. 
Furthermore, all other lines have to lie below $(\tilde a,0)$ (if one of the 
other lines $p_i$ lies well above that point, then the topological density 
becomes to a good approximation static). As the slopes of the $p_i(x_1), 
i=1,\dots,n$ are between $0$ and $1$, these graphs are never dominant once they 
are below $(\tilde a,0)$. This means that only one transition point occurs. 
Hence time-dependence of the instanton appears iff the topological charge is 
concentrated in one lump (which can be thought of as all constituents merged, 
cf.\ Fig.~\ref{fig:abs_v2_3_-1_-2}).

The case of non-distinct $\mu$'s can be understood by considering the limit 
$\mu_{j}\rightarrow\mu_{j-1}$ for some $j$'s. Then the corresponding 
constituent becomes flatter and broader, in the limit it will be invisible and 
`massless' (i.e., without topological charge/action; this has been `eaten' by 
the constituent $j+1$). In this spirit we also recover the periodic 
solution\footnote{In Yang-Mills theories this amounts to the Harrington-Shepard 
caloron \cite{harrington:78}.} with $\mu_0=\mu_1=\dots=\mu_n=0$: The 
resulting topological density then consists of only one constituent with unit 
charge, which can, but does not have to be time-dependent (depending on where 
the invisible, massless constituents are localized, see also the first row in 
Fig.~1 of \cite{Bruckmann:2007zh}). This can be demonstrated by means of 
Fig.~\ref{fig:abs_v2_-5_1_4}, i.e., based on the case of well-separated 
constituents: The limit is taken by sending the slopes of the graphs of $p_1$ 
and $p_2$ to $0$. If all positions $a_i$ are kept constant (by adjusting the 
$\lambda$-parameters), then $p_{1,2}\rightarrow 0$ (as functions) and the 
resulting topological density of the periodic solution is equivalent to the 
case of all constituents merged in the twisted model, cf.\ 
Fig.~\ref{fig:abs_v2_3_-1_-2} and Fig.~\ref{fig:density_1-instanton_cp2} (lower 
right panel). If the limit is taken with all $\lambda$-parameters kept constant 
(i.e., by sending the positions $a_{1,2}$ to $-\infty$), then 
$p_{1,2}\rightarrow2\ln\lambda_{1,2}$ can lie well above $0$ in the limit and 
the topological density remains static, though it consists of only one unit 
charged constituent.

Finally, we want to mention the possibility of generating solutions with 
topological charge less than $1$ from these instantons. Technically one has to 
avoid the asymptotics $\ln \abstx{v}\approx 4\pi x_1$ for large $x_1$ (we want 
to stay in the gauge where $\kappa_\text{min}=0$ and hence keep the asymptotics 
for small $x_1$). Hence, if the corresponding parameter 
$\lambda_{n+1}=\abstx{b^{(1)}_0}$ is vanishing (in the Fourier ansatz there is 
no integer phase $\ee^{2\pi z}$), then the total topological charge of the 
configuration is less than $1$. A phase with $\kappa_\text{max}<1$ then gives 
the total topological charge (i.e., governs the asymptotics for large $x_1$). 
Also these configurations consist of constituents with the same formulae for 
locations, sizes and charges. The number of constituents varies from $n$ down 
to $1$, depending on how many of the remaining parameters $\lambda$ are zero 
(in the graph the corresponding lines and intersection points are missing).

Interestingly, all these configurations have in common that the constituents in 
them are \emph{ordered along the noncompact direction}. This has already been 
observed in \cite{Bruckmann:2007zh} and substantiated by topological 
considerations. Here it can best be understood from 
Eq.~\eqref{eqn:const_positions}. The fractional charge of the $i$th 
constituent, $\mu_i-\mu_{i-1}$, is fixed by the twist in the boundary 
condition. These charges can be realized in isolation only if the ordering of 
their positions, $a_1\ll \dots\ll a_{n+1}$, applies. If some $a_i$ do not obey 
this ordering, then constituents emerge with the sum of the individual 
fractions as their topological charge. In other words, `pulling a constituent 
through a neighboring one' results not in a different ordering but in joining 
the constituents to a bigger one, cf.\ Fig.~\ref{fig:density_1-instanton_cp2} 
(upper and lower right panels).

Notice that by giving up our choice $\kappa_\text{min}=0$ we can
rearrange the constituents cyclically;  
this can become relevant on the lattice, where $x_1$ is of course periodic.

This ordering is of course related to the selfduality which dictates that all 
solutions are functions of $x_1+\iu x_2$; antiselfdual solutions will have the 
opposite ordering. We therefore believe that this phenomenon is particular to 
$1$+$1$ Euclidean dimensions.


\subsection[k-instanton sector]{$k$-instanton sector}\label{subsec:k-instanton}

\begin{figure*}[tb]
\centering
\includegraphics{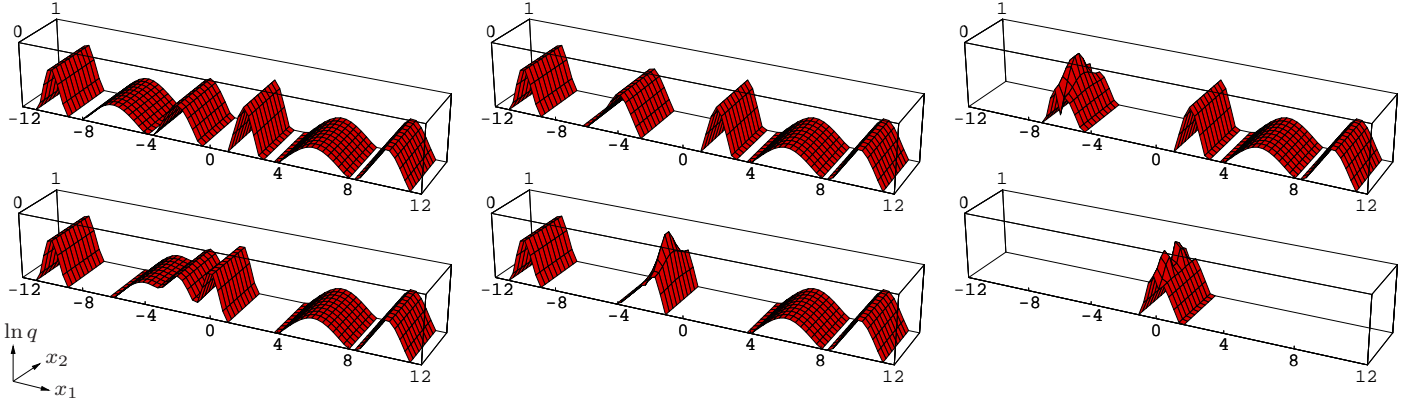}
\caption{(Color online.) Logarithm of the topological density for the 
charge-$2$ instanton of $\cp{2}$, with non-symmetric 
constituents, $\mu_1=0.55,\,\mu_2-\mu_1=0.15,\,1-\mu_2=0.3$. The 
positions of the constituents from left to right are $(a_1, a_2, a_3, a_4, a_5, 
a_6) = (-10,-6,-2,2,6,10)$, $(-10,-4,-4,2,6,10)$, $(-6,-6,-6,2,6,10)$ (first 
line) and $(-10,-4,-2,0,6,10)$, $(-10,-2,-2,-2,6,10)$, $(0,0,0,0,0,0)$ (second 
line).}\label{fig:density_2-instanton_cp2}
\end{figure*}

\noindent
The general twisted instanton solution \eqref{pinst}. 
\eqref{eq:general_twisted_instanton_solution} with integer-valued topological 
charge $k$ corresponds to the norm
\begin{multline}\label{ress1}
\abs{v(z)}^2 = \sum_{i=0}^{k(n+1)} \ee^{p_i(x_1)} + 2\sum_{s=1}^k 
\sum_{i=0}^{(k-s)(n+1)} \ee^{\tilde p_i^{(s)}(x_1)} \\
\times\cos\left(2\pi s x_2 + \varphi_{i+s(n+1)}-\varphi_i\right),
\end{multline}
where we introduced
\begin{equation}\label{andiandi}\begin{split}
p_i(x_1) & = 4\pi\mu_i x_1 + 2\ln \lambda_i,  \\
\tilde p_i^{(s)}(x_1) & = 2\pi (2\mu_i+s)x_1 + 
\ln\left(\lambda_i\lambda_{i+s(n+1)}\right)  \\
&= \frac{1}{2}\left[p_i(x_1) + p_{i+s(n+1)}(x_1)\right].
\end{split}\end{equation}
We encoded the two indices of $b^{(s)}_j$ into one, $i=s(n+1)+j$, and defined
\begin {equation}
 \label{eqn:lambda_omega}
\lambda_i=\bigl|b^{(s)}_j\bigr|,\quad \varphi_i = 
\arg\bigl(b^{(s)}_j\bigr),\quad \mu_i=\mu_j+s,
\end{equation}
where $s=0,\dots,k$, $j=0,\dots,n$. To arrive at (\ref{ress1}) we transformed 
the constant vector $b^{(k)}$ in the $0$-direction. Similarly as for the 
one-instanton solution we conclude that the constituents are localized at the 
transition points of the piecewise linear function 
$\abs{v(z)}^2$. The topological 
density thus splits into at most $k(n+1)$ constituents. Well-separated 
constituents are static and exponentially localized at 
$a_i,\,i=1,\dots,k(n+1)$, given in an analogous manner as in 
the $1$-instanton case, cf.\ Eq.~\eqref{eqn:const_positions}. The constituents 
carry the fractional charge $\mu_i-\mu_{i-1}$ and from the periodicity of 
the $\mu$'s in Eq.~\eqref{eqn:lambda_omega} follows that they are again 
ordered, see also Fig.~\ref{fig:density_2-instanton_cp2}, upper left panel.

Again we have
\begin{equation}
\tilde p^{(s)}_i (x_1) \leq \max\left\{p_i(x_1),p_{i+s(n+1)}(x_1)\right\},
\end{equation}
since the three graphs of $p_i$, $p_{i+s(n+1)}$ and $\tilde p^{(s)}_i$ 
all intersect at
\begin{equation}
\tilde a^{(s)}_i= - \frac{ \ln \left(\lambda_{i+s(n+1)}/\lambda_i\right)}{2\pi 
s}.
\end{equation}
Therefore the time-dependent term containing $\tilde p_i^{(s)}$ only 
contributes to  the sum in (\ref{ress1}) if the $s(n+1)$ constituents at the 
positions $a_{i+1},\dots,a_{i+s(n+1)}$ merge to one constituent with integer 
charge $s$. The integer $s$ thus determines the maximal frequency of the 
emerging constituent measured in units of the smallest possible frequency. We 
illustrated this behavior in Fig.~\ref{fig:density_2-instanton_cp2} for 
$2$-instanton solutions of the $\cp{2}$ model. Note that the 
freedom $\varphi_i$ of choosing the complex phase of the parameters $b_j^{(s)}$ enters 
only as shifts in the time dependence, Eq.~\eqref{ress1}.


\subsection{Cooling of lattice data}

\begin{figure*}[tb]
\centering
\includegraphics{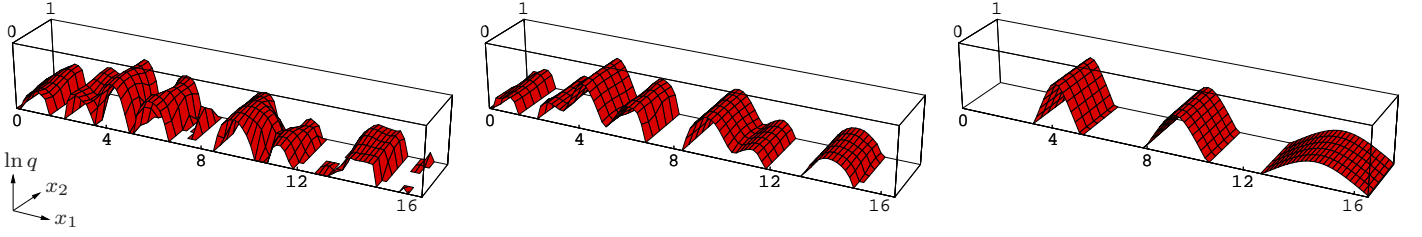}
\caption{(Color online.) The cooling procedure is applied to a $\cp{2}$ configuration of 
the Monte-Carlo ensemble with twists 
$\mu_1=0.15,\,\mu_2-\mu_1=0.5,\,1-\mu_2=0.35$. Three stable 
constituents emerge after several cooling sweeps ($10$, $25$, $500$ from left 
to right). Here only the positive part of $q$ is 
shown.}\label{fig:density-lattice}
\end{figure*}

\noindent 
Charge-one instantons of the $\cp{2}$ model containing up to the 
maximal number of three constituents are reproduced with a cooling of our 
lattice data. The simulations were performed on a $6\times 100$ 
(temporal$\times$spatial) lattice at coupling $g^{-2}=2$. In the spatial 
direction periodic boundary conditions are imposed whereas in the temporal 
direction the $v_j$ are twisted with prescribed $\mu_j$. Then a particular 
configuration is cooled. During the cooling procedure configurations with 
$\abs{k}=1$ and two or more well separated constituents are fairly stable even 
with this type of unimproved cooling (at least up to $10^5$ cooling sweeps).  
We also observe the typical annihilation of selfdual and antiselfdual constituents.
Only a small fraction of configurations is cooled to a state with \emph{three} 
clearly separated constituents. One of these is shown in 
Fig.~\ref{fig:density-lattice} at three different cooling stages. More often we 
end up with only \emph{two} constituents. These results indicate that in a 
dynamical simulation topological objects with fractional charge (given by the 
twist parameters $\mu_j$) may be as relevant as they are in Yang-Mills 
theories.


\section{Zero modes of the Dirac operator}
\subsection{Minimal coupling to fermions}
\noindent
We extend the bosonic {\cpn} model by introducing a massless Dirac fermion 
$\psi$, for the time being minimally coupled, such that the action has the form
\begin{equation}
S = \int \mathrm d^2x \left[(D_\mu u)^\dagger D_\mu u + \iu\bar\psi \gamma^\mu 
D_\mu \psi\right].
\end{equation}
We shall use the chiral representation of the $\gamma$-matrices for which 
$\gamma_*=-\iu\gamma^1\gamma^2$ is diagonal,
\begin{equation}
\gamma^1 = \left(\begin{matrix}0&1\\1&0\end{matrix}\right),\quad \gamma^2 
=\left(\begin{matrix}0&\iu\\-\iu&0\end{matrix}\right),\quad \gamma_* = 
\left(\begin{matrix}-1&0\\0&1\end{matrix}\right).
\end{equation} 
Splitting the Dirac spinor into chiral components $\psi = (\varphi,\chi)^T$ the 
Dirac equation in the background of a self-dual configuration splits into the 
Weyl equations
\begin{equation}
\begin{split}
\left(\partial-u^\dagger\partial u\right)\varphi&=\abs{v}\, 
\partial\left(\abs{v}^{-1}\varphi\right)=0 ,\\
\left(\bar\partial -u^\dagger \bar\partial 
u\right)\chi&=\abs{v}^{-1}\,\bar\partial \left(\abs{v} \chi\right) =0,
\end{split}
\end{equation}
where $\partial $ and $\bar\partial$ denote the derivatives with respect to the 
complex coordinates $z=x_1+\iu x_2$ and $\bar z=x_1 - \iu x_2$. It follows that 
the zero modes have the form
\begin{equation}
\varphi(x) = f(\bar z) \abs{v} \quad\text{and}\quad \chi(x) =  
\frac{g(z)}{\abs{v}}
\end{equation}
with (anti-)holomorphic functions $f(\bar z)$  and $g(z)$. Similarly as for the 
bosonic fields we impose quasi-periodic boundary conditions for the fermi field,
\begin{equation} 
\psi(x_1,x_2+1) = \ee^{2\pi\iu\zeta} \psi(x_1,x_2) \quad \text{with} \quad 
\zeta\in[0,1).
\end{equation}
The functions $f,g$ can be expanded in Fourier series,
\begin{equation}\label{zero_mode}
\begin{split}
\varphi(x) & = \sum_{s=-\infty}^{\infty} \alpha^{(s)} \abs{v} \ee^{2\pi 
(s+\zeta) \bar z},\\
\chi(x) & = \sum_{s=-\infty}^{\infty} \beta^{(s)} \ee^{2\pi (s+\zeta) 
z}/\abs{v}.
\end{split}
\end{equation}
\begin{figure}[tb]
\centering\includegraphics{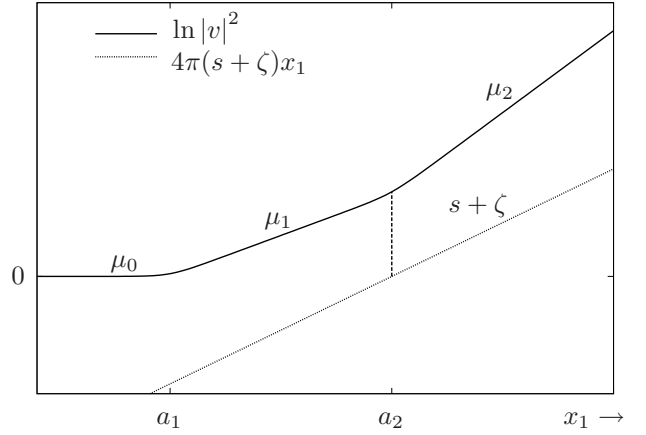}
\caption{Piecewise linear function $\ln\abs{v}^2$ and linear function $4\pi
(s+\zeta) x_1$ for the example of $s+\zeta \in (\mu_1,\mu_2)$. For
$x_1<a_1$ the slope of $\ln\abs{v}^2$ is $4\pi\mu_0=0$, for $a_1<x_1<a_2$ it
is $4\pi\mu_1$, for $a_2<x_1$ it is $4\pi\mu_2$. The vertical distance 
between the graphs is minimal at $a_2$ where the zero mode is 
localized.}\label{fig:illustration_zero-mode}
\end{figure}
These modes are only square integrable on the cylinder if the coefficients 
$\alpha^{(s)},\beta^{(s)}$ and the twist parameter $\zeta$ fulfill certain 
constraints. Recall that the asymptotic behavior of the general 
solution with charge $Q=\kappa_\text{max}$ (we set 
$\kappa_\text{min}=0$) is
\begin{equation}
\lim_{x_1\rightarrow-\infty}\abs{v} = 1 \quad \text{and}\quad 
\lim_{x_1\rightarrow\infty}\abs{v}\propto \ee^{2\pi Q x_1}.
\end{equation}
Therefore there are no normalizable left-handed zero modes 
$\varphi$.\footnote{For anti-instantons with negative topological charge the 
left-handed modes become normalizable.}  The quantum number $s$ of the 
right-handed zero modes is constrained by $0 < s+\zeta < Q$.

This immediately leads to the \emph{index theorem} for right-han\-ded zero modes. 
For integer topological charge we have to distinguish two cases for the 
fermionic phases $\zeta$:
\begin{equation}
\begin{split}
\zeta = 0 : & \quad Q-1 \text{ zero modes,}\\
\zeta \in (0,1) : & \quad Q \text{ zero modes.}
\end{split}
\end{equation}
For fractional topological charge we introduce the floor function $\lfloor 
\, \cdot \, \rfloor\colon \mathbb R \rightarrow \mathbb Z$ and the fractional part $\{ \, \cdot \, 
\}\colon \mathbb R \rightarrow [0,1)$ such that $Q=\left\lfloor Q \right\rfloor + \left\{ Q \right\}$ 
and obtain
\begin{equation}
\begin{split}
\zeta = 0 : & \quad \left\lfloor Q \right\rfloor \text{ zero modes,}\\
\zeta \in \left(0,\left\{ Q  \right\}\right)  : & \quad \left\lfloor Q 
\right\rfloor+1 \text{ zero modes,}\\
\zeta \in \left[\left\{ Q \right\},1\right)  : & \quad \left\lfloor Q 
\right\rfloor \text{ zero modes.}
\end{split}
\end{equation}

\begin{figure*}[tb] 
\centering
\includegraphics{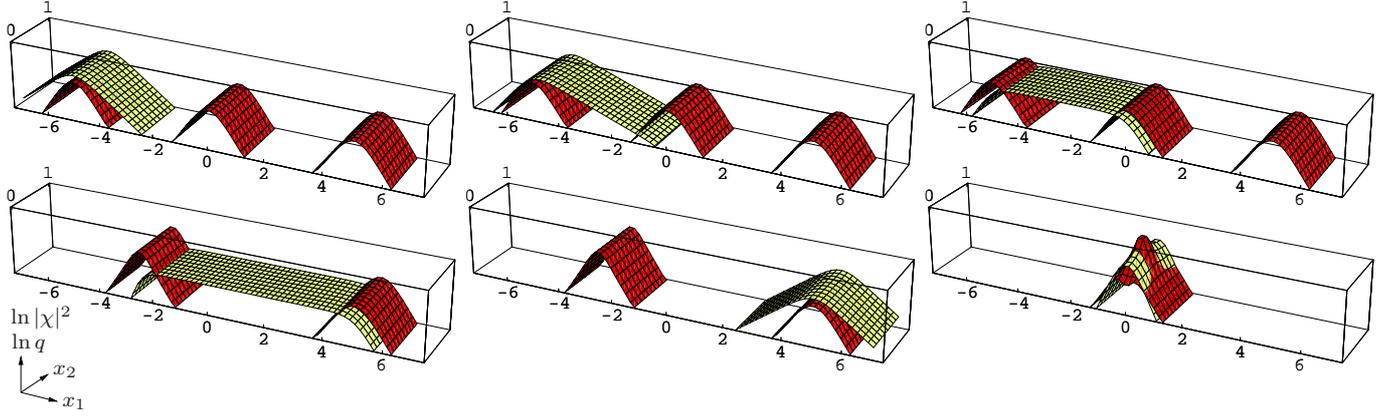} 
\caption{(Color online.) Minimally coupled fermionic zero modes (yellow/light) 
in the background of $1$-instanton constituents (red/dark) of the $\cp{2}$
model with symmetric constituents, 
$\mu_1=\mu_2-\mu_1=1-\mu_2=1/3$. In the first 
line the constituents are well-separated at $(a_1,a_2,a_3) = (-5,0,5)$, in the 
second line from left to right they are at $(-2.5,-2.5,5), (-2.5,-2.5,5), 
(0,0,0)$, i.e., two resp.\ three constituents have merged. The fermionic twist 
is $\zeta=1/6, 1/4, 1/3$ (first line) and $2/3, 5/6,1/2$ (second 
line).}\label{fig:zero_modes_1-instanton}
\end{figure*}

Let us further investigate the localization properties of the (right-handed) 
zero modes in the background of the instanton with integer charge $Q=k$. They 
have the explicit form
\begin{equation}\label{eq:abs-psi2}
\bigl|\chi^{(s)}(x)\bigr|^2= \ee^{4\pi(s+\zeta) x_1-\ln \abs{v}^2}, \quad s=0,\dots,k-1,
\end{equation}
with $\abs{v}^2$ from \eqref{ress1}. It is helpful to first consider 
well-separated constituents for which $\ln \abs{v}^2$ becomes time-independent 
and piecewise linear,
\begin{equation}
\ln \abs{v(z)}^2\approx \left\{p_i(x_1)=4\pi \mu_i 
x_1+2\ln\lambda_i\,\vert\, a_{i}< x_1< a_{i+1}\right\},
\end{equation}
as described in Sec.~\ref{subsec:k-instanton}. Clearly, the zero mode has 
maximal amplitude at points where $\ln \abs{v}^2-4\pi (s+\zeta) x_1$ is 
minimal. At these $x_1$ the vertical distance between the graphs of the 
approximately piecewise linear function $\ln \abs{v}^2$ and the linear function 
$4\pi(s+\zeta) x_1$ is minimal. For $s+\zeta$ in the interval 
$(\mu_{i-1},\mu_i)$ the minimum is at $x_1=a_i$ where the graphs of 
$p_{i-1}$ and $p_i$ intersect. The situation is depicted in 
Fig.~\ref{fig:illustration_zero-mode}.

Hence, for generic values of $\zeta$ the zero mode is \emph{localized at one 
constituent}. The profile of the zero mode is symmetric about the constituent 
$i$ for the particular value $s+\zeta=\frac{1}{2}(\mu_{i-1}+\mu_i)$,
\begin{equation} 
\bigl|\chi^{(s)}(x)\bigr|^2\propto 
\frac{1}{\cosh\left[2\pi(\mu_i-\mu_{i-1})(x_1-a_i)\right]}.
\end{equation}

Interestingly, the profile is almost constant between the $i$th and $(i+1)$th 
constituent for $\zeta=\mu_i$. These `bridges' can be understood by the fact 
that in this region the graphs of $\ln\abs{v}^2$ and $4\pi(s+\zeta) x_1$ are 
parallel (up to exponentially small corrections) at these values of $s$ and the 
fermionic phase $\zeta$.

Altogether the zero modes walk along the ordered set of constituents when 
changing the fermionic phase $\zeta$. With phases at the bounds $\zeta=0$ 
resp.\ $\zeta=1$ (or $\zeta=\{Q\}$ for configurations with fractional charge) 
the zero modes become constants asymptotically. In other words, these zero 
modes have `bridges' coming from $-\infty$ resp.\ reaching out to $+\infty$ and 
hence are not normalizable.

Similar arguments apply if two or more constituents merge. A few examples are 
given in Fig.~\ref{fig:zero_modes_1-instanton}.


\subsection{Zero modes on the lattice}
\begin{figure*}[tb]
\centering
\includegraphics{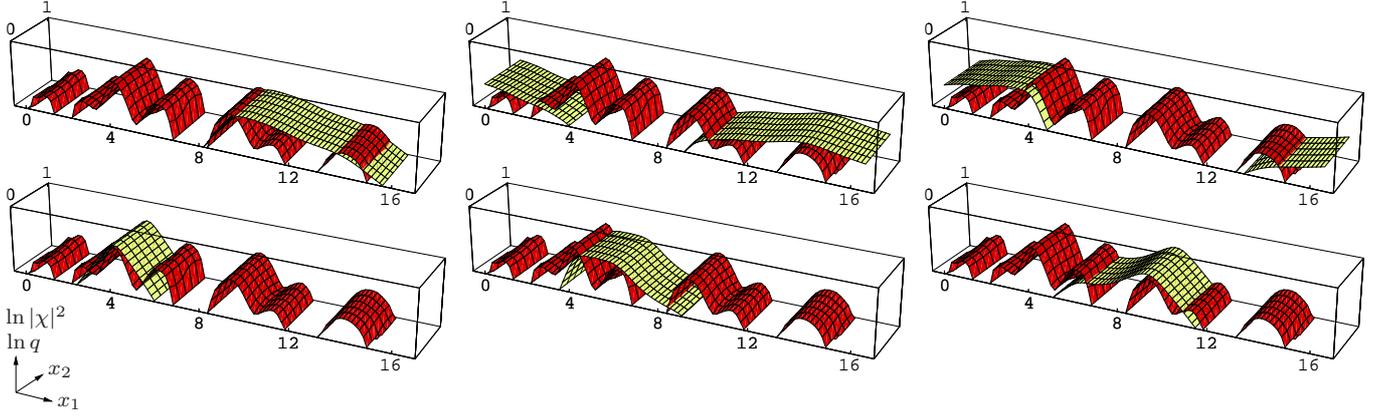} 
\caption{(Color online.) Fermionic zero modes of the overlap operator for the 
$\cp{2}$ model with twists 
$\mu_1=0.15,\,\mu_2-\mu_1=0.5,\,1-\mu_2=0.35$ after application of 
$25$ cooling sweeps (compare to 
Fig.~\ref{fig:density-lattice}, middle panel). The fermionic twist is $\zeta=0,
 0.075, 0.15$ (first line) and $\zeta=0.4, 0.65, 0.825$ (second 
line).}\label{fig:zeromodes-lattice}
\end{figure*}

\noindent
Using the overlap operator with quasi-periodic boundary conditions for the 
$U(1)$ gauge field we are able to analyze its zero modes 
\cite{Niedermayer:1998bi} in the background given by the cooled lattice 
configurations. Even for moderately cooled configurations do the zero modes 
reflect the position of the fully cooled instanton constituents for specific 
boundary conditions (see Fig.~\ref{fig:zeromodes-lattice}). Therefore the 
lattice results are in full agreement with the analytical results and in 
addition the fermionic zero modes are excellent tracers for instanton 
constituents for mildly cooled configurations.

\subsection{Supersymmetric coupling to fermions}
\noindent
The supersymmetric {\cpn}  model \cite{DAdda:1978kp, Witten:1978bc} contains 
$n+1$ Dirac fermion fields $\psi_j$, $j=0,\dots,n$, in addition to the complex 
scalar fields $u_j$. Its action is
\begin{multline}
S = \int \mathrm d^2x \biggl\{\left(D_\mu u\right)^\dagger \left(D_\mu u\right) 
+ \iu\bar\psi \gamma^\mu D_\mu \psi  \\
+ \frac{1}{4}\left[\left(\bar\psi\psi\right)^2 - 
\left(\bar\psi\gamma_*\psi\right)^2 - \left(\bar\psi\gamma^\mu\psi\right) 
\left(\bar\psi\gamma_\mu\psi\right)\right]\biggr\},
\end{multline}
where $u$ and $\psi$ are constrained by
\begin{equation}
u^\dagger  u = 1, \qquad u^\dagger\psi = \bar\psi u = 0.
\end{equation}
Introducing Weyl spinors, $\psi=(\varphi,\chi)^T$, the model is invariant under 
the on-shell $\mathcal{N}=(2,2)$ supersymmetry transformations
\begin{equation}\begin{split}
\delta u & = \varepsilon_1\varphi - \varepsilon_2\chi, \\
\delta\varphi & =  +2\iu\bar\varepsilon_1\bar D u - \bar\varepsilon_1 
(\bar\varphi\varphi)u + \bar\varepsilon_2(\bar\chi\varphi) u,\\
\delta\chi & = - 2\iu\bar\varepsilon_2D u + \bar\varepsilon_2(\bar\chi\chi)u - 
\bar\varepsilon_1(\bar\varphi\chi) u,
\end{split}\end{equation}
with the covariant derivative
\begin{equation}
D = \partial - u^\dagger \partial u\quad \text{and}\quad \bar 
D=\bar\partial-u^\dagger\bar\partial u
\end{equation}
and anticommuting parameters $\varepsilon_ {1,2}$ satisfying $\varepsilon_2^* 
=\bar\varepsilon_1$ and $\varepsilon_1^* =\bar\varepsilon_2$. Both spinors 
$\varphi$ and $\chi$ have $n+1$ components.

\begin{figure}[t]
\centering\includegraphics{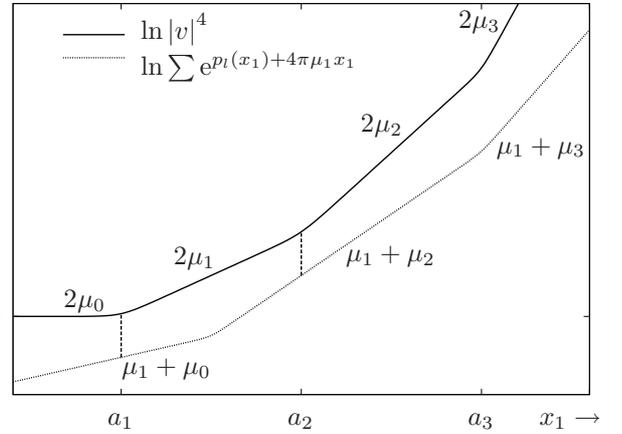}
\caption{Logarithm of denominator and numerator of $\abs{\chi^{(1)}}^2$ in
\eqref{eq:chi(i)}. The zero mode has two maxima of equal amplitude at $a_1$ and
$a_2$.}\label{fig:illustration_zero-mode_susy}
\end{figure}

\begin{figure*}[tb]
\centering
\includegraphics{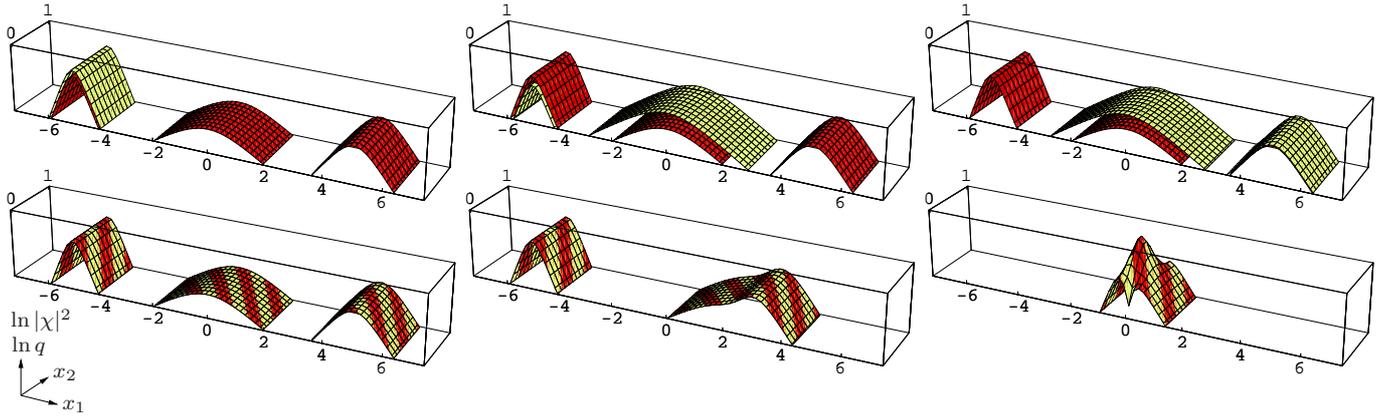}
\caption{(Color online.) Supersymmetric coupled fermionic zero modes
(yellow/light) in the background of $1$-instanton constituents (red/dark) in the
$\cp{2}$ model with twist parameters $\mu_1=0.55$,
$\mu_2-\mu_1=0.15$, $1-\mu_2=0.3$. In 
the first line the zero modes $\chi^{(0)}$, $\chi^{(1)}$ and $\chi^{(2)}$ (from
left to right) with instanton locations $(a_{1}, a_{2}, a_{3}) = (-5,0,5)$. In
the second line the half-BPS state $\chi_\text{inst}$ with instanton locations
$(-5,0,5), (-5,2,3), (0,0,0)$ (from left to right).}\label{fig:zero_modes_susy}
\end{figure*}

The \emph{linearized} Dirac equation in an external $u$-field splits into two 
Weyl equations,
\begin{equation}
\left(1-uu^\dagger\right)D\varphi=0 \quad\text{and}\quad \left(1 - u 
u^\dagger\right) \bar D\chi = 0.
\end{equation}
For an instanton background with $u=v(z)/\abstx{v}$ the Weyl equations simplify 
to
\begin{equation}
\left(1-P_v\right)\, \partial\left( \abs{v}^{-1}\varphi\right)= 
\left(1-P_v\right)\,\bar\partial\left(\abs{v}\chi\right)=0,
\end{equation}
where $P_v$ projects onto the holomorphic $v(z)$,
\begin{equation}
P_v=uu^\dagger=\frac{v v^\dagger}{\abs{v}^2}.
\end{equation}
It follows that a left-handed solution reads
\begin{equation}
\varphi(x) =\abs{v} f(\bar z),
\end{equation}
where $f(\bar z)$ is an arbitrary vector of anti-holomorphic functions 
orthogonal to $v$. None of these solutions is normalizable. With the help of 
$P_v\bar\partial P_v=\bar\partial P_v$ one shows that a right-handed solution 
has the form
\begin{equation}\label{asol}
\chi(x) = \frac{1}{\abs{v}}\left(1 - P_v\right) g(z),
\end{equation}
where $g(z)$ is a vector of holomorphic functions. In order not to break 
supersymmetry $g$ must fulfill the same boundary conditions as the instanton 
solution $v$. Therefore, the choice of fermionic twists is very limited here. 
Each function $g$ can be constructed by linear combination of the basis 
elements $\{g^{(j,s)}\}$ defined by
\begin{equation}
g^{(j,s)}(z) = \ee ^{2\pi (s+\mu_j) z}e_j, \quad j=0,\dots,n, \quad s\in 
\mathbb Z,
\end{equation}
where $e_j$ is the unit vector pointing in direction $j$. For the corresponding 
zero modes the squared norm is
\begin{equation}
\bigl|\chi^{(j,s)}\bigr|^2 = 
\frac{\ee^{4\pi(s+\mu_j)x_1}}{\abs{v}^4}\sum_{l\neq j}^n \abs{v_l}^2.
\end{equation}
Normalizability of the zero mode in the $k$-instanton background requires
\begin{equation}
s = \begin{cases} 0,1,\dots,k & \text{for }j=0,\\
0,1,\dots,k-1 & \text{for }j=1,\dots,n.
\end{cases}
\end{equation}
In the case of well-separated instanton constituents we can write
\begin{equation}\label{eq:chi(i)}
\bigl|\chi^{(i)}\bigr|^2 \approx \frac{1}{\abs{v}^4}\sum_{l \bmod{(n+1)}\neq 
i}^{k(n+1)} \ee ^{p_l(x_1) + 4\pi\mu_i x_1},
\end{equation}
where we introduced $\chi^{(i)} = \chi^{(j,s)}$ for $i=s(n+1)+j$. The linear 
functions $p_l(x_1)$ are given in (\ref{andiandi}). Again the maximum of 
$\abstx{\chi^{(i)}}$ is easily found by considering the graphs of the linear 
functions $2p_l(x_1)$ and $p_l(x_1) + 4\pi\mu_i x_1$. In a logarithmic plot 
both the numerator and denominator of $\abstx{\chi^{(i)}}$ are piecewise 
linear. For $x_1<a_{i}$ the slope of the numerator is larger and for 
$x_1>a_{i+1}$ the slope of the denominator is larger. This is illustrated in 
Fig.~\ref{fig:illustration_zero-mode_susy}.
Simple geometric arguments about these graphs reveal, that the zero modes 
$\chi^{(i)}$ with $0<i<k(n+1)$ split into two constituents located at $a_i$ and 
$a_{i+1}$, which have the same amplitude, but decay with different lengths. The 
zero modes $\chi^{(0)}$ and $\chi^{(k(n+1))}$ have only one maximum at $a_1$ 
and $a_{k(n+1)}$, respectively. Some examples are plotted in 
Fig.~\ref{fig:zero_modes_susy}. The general right-handed zero mode has the form
\begin{equation} \label{eq:chi-general}
\chi = \sum_{i=0}^{k(n+1)} \beta_i \chi^{(i)}.
\end{equation}
Its (squared) norm splits into $k(n+1)$ or less constituents. They have the 
same analytic form $\propto \cosh^{-2} \left[2\pi 
Q_{\text{const},i}(x_1-a_i)\right]$ and are located at the same positions $a_i$ 
as the instanton constituents.

There exists always a particular zero mode, whose (squared) norm is 
proportional to the topological density
\begin{equation}
q =\frac{1}{4\pi}\Delta\ln \abs{v}^2= \frac{1}{\pi}\frac{\abs{v}^2\abs{\partial 
v}^2 - \abs{v^\dagger \partial v}^2}{\abs{v}^4}.
\end{equation}
Namely, since the squared norm of $\chi$ in (\ref{asol}) is
\begin{equation}
\abs{\chi}^2 = \frac{\abs{v}^2 \abs{g}^2 - \abs{v^\dagger g}^2}{\abs{v}^4},
\end{equation}
we obtain the exact relation
\begin{equation}
\abs{\chi(x)}^2 =\pi q(x)
\end{equation}
for the zero mode with $g =\partial v$, which means that $\beta_i\propto 
\mu_i\lambda_i$ in Eq.~\eqref{eq:chi-general}.

The occurrence of this particular zero mode can also be understood as follows: 
Any instanton background breaks half of the supersymmetry, namely the one 
generated by the parameter $\bar\varepsilon_2$. If we transform the 
configuration $u_\text{inst}= v(z)/\abs{v}$, $\psi_\text{inst}=0$ with the 
broken symmetry then $\delta\psi_\text{inst}$ is inevitably a zero mode of the 
Dirac operator for the action is invariant under the supersymmetry 
transformation. This way we obtain a non-vanishing right-handed zero mode
\begin{equation}
\delta\chi_\text{inst} = -2\iu \bar\varepsilon_2 D u \propto \frac{1}{\abs{v}} 
\left(1 - P_v\right)\partial v,
\end{equation}
that is a zero mode with $g\propto \partial v$. Except for irrelevant 
prefactors, this is exactly the zero mode whose squared norm is equal to the 
topological density of the instanton.


\section{Conclusions}
\noindent
In the present work we constructed and analyzed the integer-charged instantons 
for twisted {\cpn} models on a cylinder. The twisted instantons with charge $k$ 
support $k(n+1)$ constituents. If these constituents are well-separated then 
they become static lumps. The fractional charges and the shapes of the 
constituents topological profile are governed by the phases in the boundary 
condition (and the scale $\beta$). The constituent positions are related to the 
collective parameters of the twisted instanton and hence free up to the demand 
that for all constituents to be pre\-sent their positions must be ordered.

Neighboring constituents can merge adding up their char\-ges. If at least 
$n+1$ constituents merge then the resulting lump becomes time-dependent. For a 
composite object containing multiples of $n+1$ constituents time-dependent 
terms with higher frequencies contribute, respectively.

Our analytic findings are in complete agreement with the corresponding 
numerical ones. The latter were obtained by cooling lattice configurations of 
the twisted model with a non-vanishing topological charge.

We determined all fermionic zero modes in the background of the twisted 
instantons. This has been achieved for minimally coupled fermions satisfying 
quasi-periodic boundary conditions in the Euclidean time direction. We found 
that, similarly as for gauge theories, the zero modes are localized at the 
positions of the constituents and that they may jump from one constituent to 
the neighboring one if the boundary conditions for the fermions are changed. 
Again we compared our analytical findings to numerical results. To that aim we 
determined the zero modes of the overlap Dirac operator for lattice 
configurations with different degrees of cooling. Again we find full agreement 
between our analytical and numerical results, in close analogy with the 
corresponding situation for $SU(N)$ Yang-Mills theories.

In the supersymmetric {\cpn} model the Dirac fermions transform according to 
the fundamental representation of the global $U(n+1)$ symmetry group. The 
linearized field equations for the $n+1$ fermion-flavors define a 
supersymmetric Dirac operator. We studied the square integrable zero modes of 
this operator and showed that they generically split into $k(n+1)$ constituents 
with maxima at the locations of the instanton constituents. There 
exists always a particular zero mode whose norm squared is equal to the 
topological charge density of the supporting instanton. This zero mode is 
generated by the half-broken supersymmetry. We did not elaborate on the 
contribution of the constituents and zero modes to the central charge of the 
$(2,2)$ SUSY algebra.

Our results are in close parallel to the corresponding findings in $SU(N)$ gauge 
theories. But since twisted instantons, their constituents and the fermionic 
zero modes in {\cpn} models are much simpler as in gauge theories our results 
may be useful to shed further light on the relevant degrees of freedom in 
strongly coupled models at finite temperature. The next natural step would be 
to include quantum fluctuations about twisted instantons to study the quantum 
corrections to the constituent picture.

In the $SU(N)$ gauge theory there is a beautiful construction of the 
constituents based on the Nahm transform \cite{Kraan:1998pm}. We believe that a 
similar construction, with Nahm transform as introduced in 
\cite{Aguado:2001xg}, could further simplify the construction of instanton 
constituents for twisted {\cpn} models.

Similar aspects of twisted {\cpn} models (such as loop groups) are discussed in
a simultaneous paper \cite{Harland:2009}.


\section*{Acknowledgments}
\noindent
We thank Michael Thies for discussions and Ulrich Theis for pointing us to the 
BPS-interpretation of one of the zero modes of the supersymmetric Dirac 
operator. W. Brendel, L. Janssen and C. Wozar thank for the support by the 
Studienstiftung des deutschen Volkes. This work has been supported by the DFG 
grants Wi 777/10-1 and BR 2872/4-1.


\bibliography{literature}

\end{document}